%% file: main.tex
\documentclass[12pt,headings=normal]{article}
\pdfoutput=1
\usepackage[utf8]{inputenc}

\usepackage[normalem]{ulem}
\usepackage{amsmath,amssymb}
\usepackage{slashed}
\usepackage{xcolor}
\usepackage{setspace}

\usepackage{graphicx}
\usepackage{cite}
\usepackage{pdflscape}
\usepackage{multirow}
\usepackage[thinlines]{easytable}
\usepackage{url}
\usepackage{braket}
\usepackage{subfigure}
\usepackage{longtable}
\usepackage{booktabs}

\input{shortcuts.tex}

\begin{document}

\begin{titlepage}

\begin{center}
\begin{spacing}{1.6}
{\LARGE 
Matter Coupling in Massive Gravity
}\end{spacing} %
\vspace{1.4cm}

\renewcommand{\thefootnote}{\fnsymbol{footnote}}
{Adam~Falkowski and  Giulia~Isabella}
\renewcommand{\thefootnote}{\arabic{footnote}}
\setcounter{footnote}{0}

\vspace*{.8cm}
\centerline{\em 
Universit\'{e} Paris-Saclay, CNRS/IN2P3, IJCLab, 91405 Orsay,
France }\vspace{1.3mm}
\vspace{.2cm}
\em{E-mail:~} falkows@ijclab.in2p3.fr, isabella@ijclab.in2p3.fr

\vspace*{.2cm}

\end{center}

\vspace*{10mm}
\begin{abstract}\noindent\normalsize
We discuss the dRGT massive gravity interacting with spin-0, spin-1/2, or spin-1 matter. The effective theory of a massive spin-2 particle coupled to matter particles is constructed directly at the amplitude level. In this setting we calculate the gravitational Compton scattering amplitudes and study their UV properties. While the Compton amplitudes generically grow with energy as $\cO(E^6)$, we identify regions of the parameter space where they are softened to $\cO(E^4)$ or even $\cO(E^3)$, which allows for a larger validity range of the effective theory. In these regions, both positivity and beyond-positivity of the forward Compton amplitudes are fulfilled, and the equivalence principle automatically emerges.

\end{abstract}

\end{titlepage}
\newpage

\renewcommand{\theequation}{\arabic{section}.\arabic{equation}}

\tableofcontents

\section{Introduction}
\label{sec:intro}
\setcounter{equation}{0}

Attempts to modify gravity at large distances have a long history. 
Perhaps the most appealing example is the class of theories where the spin-2 carrier of the gravitational force  - the graviton - has a tiny mass $m$~\cite{FP}.   
As this makes gravity a finite-range force, $m$ cannot be much larger than the inverse Hubble length~\cite{deRham:2016nuf}.
Construction of consistent and phenomenologically viable theories of this kind encounters many practical difficulties.
One of them is the rapid growth of graviton scattering amplitudes for energies $E \gg m$.
As a result, any known effective field theory (EFT) of a massive graviton  hits the strong coupling at $E \sim \Lambda_s$ where  $\Lambda_s^{-1}$ is a macroscopic distance scale. 
As the cutoff scale $\Lambda$ of the EFT must satisfy $\Lambda \lesssim \Lambda_s$, this severely limits the possible validity range of massive gravity theories. 
 
From the phenomenological point of view it is beneficial to postpone the onset of strong coupling as much as possible, thus extending the predictive power of the EFT. 
This can be achieved by choosing the interactions of the graviton so as to make the scattering amplitudes softer.\footnote{Notice that we use the term {\em soft} in the opposite sense as e.g. in Ref.~\cite{Bellazzini:2016xrt}. 
In this paper, for amplitudes growing as $\cM \sim E^n$ for $m \ll E \ll \Lambda$, softer means a {\em smaller} power $n$. } 
For the $2 \to 2$ graviton self-scattering amplitude one can arrange for 
$\cM(G G\to GG) \sim (E/\Lambda_3)^6$~\cite{ArkaniHamed:2002sp}, where $\Lambda_3  = (m^2 \mpl)^{1/3}$ and $\mpl \approx 2.4 \times 10^{18}$~GeV.  
The concrete, non-linear, ghost-free realization of this scenario is the {\em dRGT gravity}~\cite{deRham:2010ik,deRham:2010kj,Hinterbichler:2011tt,deRham:2014zqa}.
This is an EFT of a single massive spin-2 particle with the strong coupling scale given by $\Lambda_3$,
also when n-point graviton amplitudes with $n > 4$ are taken into account.  

In this paper we discuss dRGT gravity coupled to matter, where the latter stands for massless or massive spin-0, spin-1/2, or spin-1 particles.  
We build the corresponding EFT directly at the level of on-shell amplitudes in the Minkowski background, without passing through fields and Lagrangians.
This is a great simplification when massive spin-2 particles are involved: one deals only with the 5 physical polarizations, while unphysical degrees of freedom (in the standard approach carried by the metric field) are never introduced into the picture.  
Consequently, calculation of amplitudes in this framework is much simpler than obtaining them through the Feynman rules from a Lagrangian. 

Our philosophy closely follows the one in Refs.~\cite{Bonifacio:2018vzv,Bonifacio:2019mgk}, where the on-shell amplitude formalism was applied to self-scattering of massive gravitons.  
Here we focus on the gravitational Compton scattering:  $\cM_c \equiv \cM(X G \to X G)$. 
We first build the on-shell 3-point $\cM(XXG)$ amplitudes describing the minimal coupling of the massive graviton to a matter particle $X$. 
They have the same form as the ones predicted by  Einsten's general relativity (GR), up to an overall multiplicative factor $c_X$.  
For $m > 0$, that factor (which can be interpreted as the coupling strength between gravity and matter) is allowed to deviate from the GR value $c_X = 1$. 
In other words, the equivalence principle is not assumed at the outset when the graviton has a mass. 
Two more ingredients are necessary to calculate tree level Compton amplitudes.
One is the 3-graviton amplitude, which is taken to be exactly the one predicted by the dRGT gravity. 
The other is a set of 4-point $XXGG$ contact terms, which can be organized into a systematic EFT expansion in $E/m$. 
The  final result depends on several free parameters: the coupling strength between gravity and matter, the Wilson coefficients of the contact terms, and one more parameter characterizing the 3-graviton amplitude in dRGT.   
We will take advantage of this parameter space  to regulate the UV properties of the Compton amplitudes.\footnote{In this paper,  {\em UV behavior} or {\em UV properties} always refer to energies $m \ll E \ll \Lambda$, that is above all particles' masses but within the validity range of the EFT. We are not concerned with the true UV properties of the amplitudes, that is for $E \to \infty$,  expect maybe for general statements like the Froissart bound~\cite{Froissart:1961ux}.}

The Compton amplitudes calculated at tree level display a number of interesting properties. 
For a generic point in the parameter space they grow with energy as $\cM_c \sim (E/\Lambda_3)^6$ for any spin of the matter particle, which is the same behavior as for graviton self-scattering amplitudes
Thus, they become strongly coupled around the same scale as $\cM(G G\to GG)$.
A priori, it is not necessary to adjust any parameters of this EFT so as to regulate the UV properties of $\cM_c$. 
It is interesting to observe, however, that in certain regions of the parameter space the  behavior is considerably  softer:  $\cM_c \sim E^4/m^2 \mpl^2$ or even $\cM_c \sim E^3/m \mpl^2$.
This is possible for any mass and spin of the matter particle
provided its coupling strength to the massive graviton has precisely the  value predicted by GR, $c_X=1$. 
That is to say, the equivalence principle in massive gravity can be restored simply be demanding a certain high-energy behavior of the gravitational Compton scattering amplitudes.   

Our paper is organized as follows. 
In \sref{drgt} we review the on-shell formulation of graviton self-interactions in dRGT gravity,  
and we take that opportunity to introduce our notation and conventions.
The main results are contained in \sref{matter} where we construct the leading interactions of the massive graviton with matter and calculate the Compton scattering amplitudes.
We write down the precise constraints on the parameter of the theory that lead to Compton amplitudes softer than $\cO(E^6)$.
Additional constraints on the parameter space  can be obtained assuming the UV completion of our EFT is local, causal, and respects Poincar\'{e} invariance.
These so-called positivity bounds are discussed in \sref{pos}, and we show that they are satisfied in the parameter region where the Compton amplitudes are softer.

\section{dRGT on shell}
\label{sec:drgt}
\setcounter{equation}{0}

In this section we review the calculation of $2\rightarrow 2$ scattering of massive gravitons in the dRGT gravity~\cite{Cheung:2016yqr,Bellazzini:2017fep,Bonifacio:2018vzv}. Much as Ref.~\cite{Bonifacio:2018vzv}, we work in the on-shell amplitude framework, without introducing the graviton field or Lagrangian.  
Instead, we first write down the most general 4-graviton amplitude consistent with Poincar\'{e} invariance, unitarity, and locality. 
This general form is constrained by requiring the UV behavior of the amplitudes to be as soft as possible. 
In the case of massive gravitons the best possible situation is $\cM \sim \cO(E^6)$~\cite{ArkaniHamed:2002sp}, which defines the dRGT gravity. 
This method not only simplifies the calculations, but also avoids all the subtleties of working with higher-spin Lagrangians. 

To be specific, the amplitude with four gravitons takes  the form 
\bea 
\label{eq:DRGT_M4graviton}
\cM(1234)  & = & -  \sum_h \left [  {\cM(12 \hat p_s^h) \cM(34 p_s^h) \over s - m^2 }
 + {\cM(13 \hat p_t^h) \cM(24 p_t^h) \over t - m^2} 
  + {\cM(14 \hat p_u^h) \cM(23 p_u^h) \over u - m^2}   \right ] 
  \nnl &+ &  C(1234) , 
\eea 
where  $1\dots 4$ label the external gravitons, 
$m$ is the graviton mass,  
$p_s \equiv p_1 + p_2$, $p_t \equiv  p_1 + p_3$, $p_u \equiv  p_1 + p_4$, 
the Mandelstam invariants are $i \equiv p_i^2$  for $i = s,t,u$, and the sum goes over polarizations of the intermediate graviton.
By default all particles in the amplitudes are incoming; if a particle is outgoing, the corresponding entry is marked by a hat. 
The first line contains the pole terms, schematically represented in \fref{self}. 
Their form is fixed by unitarity, which requires that the residue of each pole is given by the product of on-shell 3-graviton amplitudes. 
Note that for massive particles the poles are separated, in the sense that a residue in one channel does not contain a pole in another channel~\cite{Arkani-Hamed:2017jhn}, unlike what happens for massless graviton scattering.  
The last term denotes 4-graviton contact terms, which are regular functions of $s,t,u$ without poles or other singularities, therefore they are not connected to 3-point amplitudes by unitarity.  
In the on-shell approach the contact terms can be adjusted so as to soften the behavior of the amplitude for $E \gg m$, where $E \sim \sqrt s$ is the characteristic energy scale of the scattering process. 
In other words, the contact terms are chosen so as to maximize the validity range of the EFT of massive gravitons.  

\begin{figure}[tb]
\centering
\begin{minipage}{.3\textwidth}
  \centering
  \vspace{-0.5cm}
  \textbf{s-channel}
  
   \vspace{0.5cm}
  \includegraphics[width=.8\linewidth]{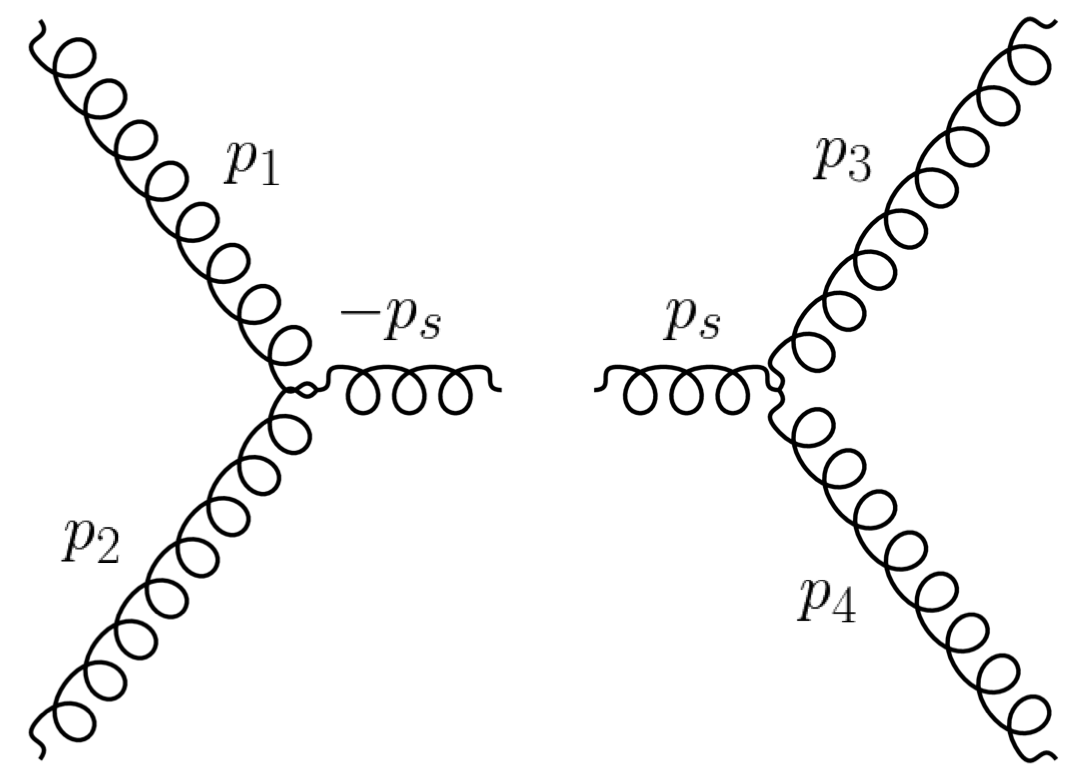}
\end{minipage}%
\begin{minipage}{.3\textwidth}
  \centering
  \vspace{-0.3cm}
  \textbf{t-channel}
  
   \vspace{0.3cm}
  \includegraphics[width=.5\linewidth]{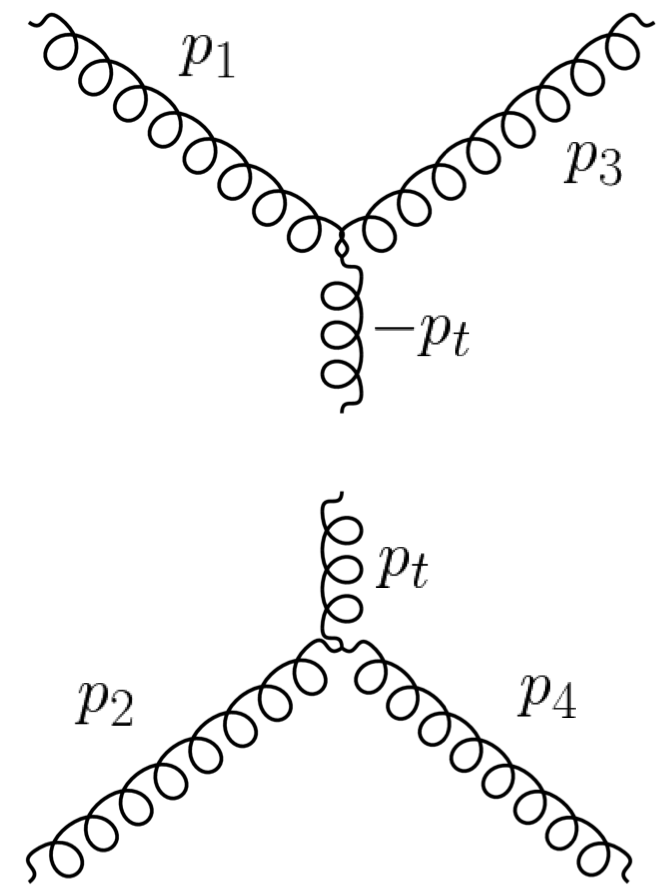}
\end{minipage}
\begin{minipage}{.3\textwidth}
  \centering
  \vspace{-0.3cm}
  \textbf{u-channel}
  
   \vspace{0.3cm}
  \includegraphics[width=.5\linewidth]{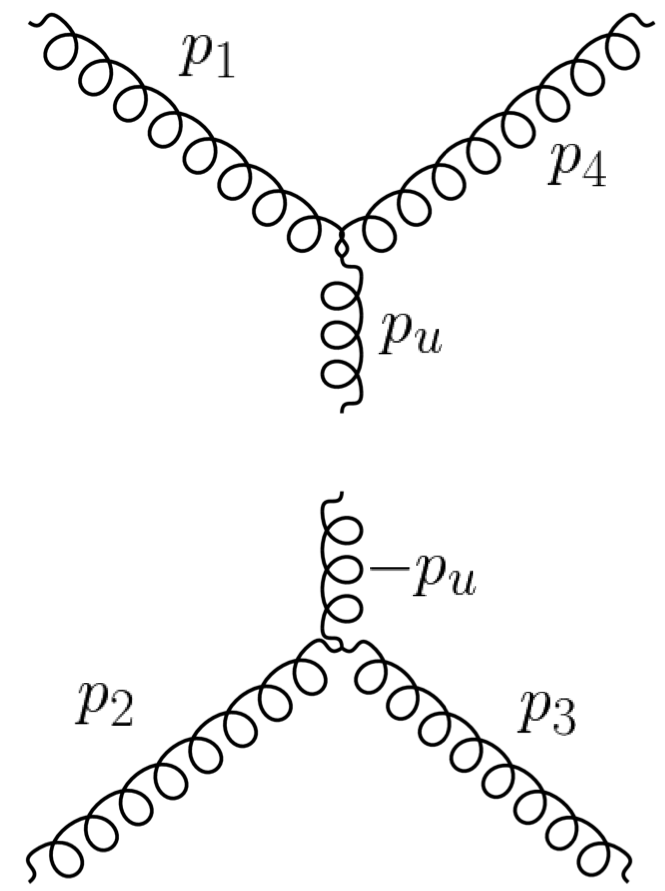}
\end{minipage}
\caption{  \label{fig:self}
Schematic representation of the pole terms in the 4-graviton amplitude in \eref{DRGT_M4graviton}. }
\end{figure}

\subsection{Polarization tensors}

In the on-shell framework, amplitudes are cast in a form that makes manifest their little group transformation properties. 
For massless particles, one works with the spinor helicity variables $\lambda$, $\tilde \lambda$, which are 2-component spinors related to the four-momenta by $p^\mu_i \sigma_\mu =\lambda \tilde \lambda$. 
The $U(1)$ little group acts on the spinors as $\lambda  \to t^{-1} \lambda$, $\tilde \lambda  \to t\tilde \lambda$.  
For massive particles the little group is $SU(2)$.  
In this case a convenient formalism~\cite{Arkani-Hamed:2017jhn} is to assign two spinor pairs  $\chi^J$, $\tilde \chi^J$ for each four-momentum, which satisfy 
$p^\mu_i \sigma_\mu = \sum_J \chi^J \tilde \chi_{ J}$ 
and are normalized as  $(\chi^J \chi_K) = \delta^J_K  m$, $(\tilde \chi_J \tilde \chi^K) =   \delta^J_K  m$. 
Here $J = 1,2$ is the $SU(2)$ little group index, which is lowered by $\epsilon_{JK}$ and raised by $\epsilon^{JK}$ antisymmetric tensors. 
For spin $S$, the appropriate little group representation is obtained by multiplying $S$ massive spinors and symmetrizing their little group indices.   
In particular, massive graviton amplitudes should contain 4 spinors $\chi^J$ or $\tilde \chi^J$  for each external graviton.
We  introduce traceless symmetric Lorentz tensors constructed out of 4 spinors:   
\beq
\label{eq:DRGT_massivespin2}
[\eps^{\mu \nu}(p)]^{JKLM} = {1 \over 2 m^2} (\chi^J \sigma^\mu \tilde \chi^K)  (\chi^L \sigma^\nu \tilde \chi^M), \qquad [\bar \eps^{\mu \nu}(p)]_{JKLM} =  {1 \over 2 m^2} (\chi_J \sigma^\mu \tilde \chi_K) (\chi_L \sigma^\nu \tilde \chi_M) , 
\eeq 
where full symmetrization of the little group indices is implicit.
These are nothing but the usual polarization tensors~\cite{Guevara:2018wpp}. 
We find it more convenient to build our amplitudes using the polarization tensors.\footnote{See Refs.~\cite{Christensen:2018zcq,Durieux:2019eor,Aoude:2019tzn,Bachu:2019ehv} for amplitudes of massive EFTs constructed out of $\chi^J$ and $\tilde \chi^J$ instead.} 
In the helicity basis, the scalar polarization corresponds to $[\eps^{\mu \nu}(p)]^{1122}$, the vector ones to  $[\eps^{\mu \nu}(p)]^{1112}$ and $[\eps^{\mu \nu}(p)]^{1222}$, and the tensor ones to $[\eps^{\mu \nu}(p)]^{1111}$ and $[\eps^{\mu \nu}(p)]^{2222}$. 
Summing a product of the polarization tensors over the little group indices one obtains the numerator of the massive graviton propagator: 
\beq
\label{eq:DRGT_numerator}
N_{\mu \nu, \alpha \beta}^p \equiv  \sum_{JKLM} [\eps_{\mu \nu}(p)]^{JKLM} [\bar \epsilon_{\alpha \beta} (p)]_{JKLM} = 
 {1 \over 2} \left ( \Pi_{\mu \alpha}  \Pi_{\nu \beta} +   \Pi_{\mu \beta}  \Pi_{\nu \alpha} \right )
 - {1 \over 3}  \Pi_{\mu \nu}  \Pi_{\alpha \beta}, 
\eeq 
where  $\Pi_{\mu \nu} = \eta_{\mu \nu} - {p_\mu p_\nu \over m^2}$. 
In the following we do not display the little group indices, and abbreviate 
$[\eps^{\mu \nu}(p_i)]^{JKLM} \equiv \epsilon_i^{\mu\nu}$. 

$N$-point graviton amplitudes can be written down in terms of Lorentz-invariant contractions $\epsilon_i$ and $p_i$, $i = 1 \dots N$, where each  $\epsilon_i$ appears exactly once. 
This automatically ensures the correct little group transformation properties.  
The operation of crossing an incoming graviton into an outgoing one amounts simply to replacing the corresponding polarization tensor with its conjugate: $\epsilon(p) \to \bar \epsilon(-p)$. 
Furthermore, working with polarization tensors makes  power counting transparent. 
Indeed, the scalar polarization of a massive graviton  can be represented by  $\epsilon_i^{\mu \nu} \sim p_i^\mu p_i^\nu/m^2$, thus in the UV each $\epsilon_i$ implicitly carries two powers of energy $E$.
Each additional momentum insertion adds another power of energy. 
This power counting will allow us to build  the ingredients of \eref{DRGT_M4graviton} - 
the 3-point amplitude and the 4-point contact terms  - in a controlled expansion in $E/m$.

\subsection{3-graviton amplitude and pole terms} 

In dRGT gravity the on-shell 3-graviton amplitude can be parametrized as 
 \beq
\label{eq:DRGT_M3drgt}
\cM(\mathbf{1} \mathbf{2} \mathbf{3})  =   
{a_0 m^2 \over \mpl}  \langle  \epsilon_1 \epsilon_2 \epsilon_3   \rangle  - 
 {1 \over 2 \mpl} \left [ 
  \langle  p_{23} \epsilon_1 p_{23} \rangle  \langle \epsilon_2 \epsilon_3 \rangle
-  2 \langle  p_{23} \epsilon_1 \epsilon_3 \epsilon_2 p_{13} \rangle   + {\rm  cyclic} \right ] , 
\eeq
where   $p_{jk} \equiv p_j - p_k$, 
and  we abbreviate the Lorentz contractions:  
$\langle \epsilon_j \epsilon_k \rangle \equiv \epsilon_j^{\mu \nu} \epsilon_k^{\mu \nu}$,  
$\langle \epsilon_j \epsilon_k \epsilon_l \rangle \equiv \epsilon_j^{\mu \nu} \epsilon_{k}^{\nu \rho} \epsilon_{l}^{\mu\rho}$,
$\langle  p_j \epsilon_l p_k \rangle \equiv p_j^\mu \epsilon_l^{\mu \nu} p_k^\nu$. 
{\em Cyclic} stands for 4 other terms obtained by cyclic permutations of the first 2 terms in the square bracket, 
so that the amplitude is Bose symmetric.
The coefficient $a_0$ of the first term  is a free parameter in this framework, related  to the commonly used parameter $c_3$ in the dRGT Lagrangian by  $a_0 = 3 (1 - 4 c_3)$.  
The second term has exactly the same form as in ordinary GR, which allows us to identify 
$\mpl = (8 \pi G)^{-1/2} \approx 2.4 \times 10^{18}$~GeV.

Given \eref{DRGT_M3drgt}, we can rewrite the 4-graviton amplitude in \eref{DRGT_M4graviton} as 
\bea 
\label{eq:DRGT_M4drgt}
\cM(1234)  & = & -   {M_{\mu \nu}(12) N_{\mu \nu, \alpha \beta}^{p_s} M_{\alpha \beta}(34) \over s - m^2 }
 - {M_{\mu \nu}(13)N_{\mu \nu, \alpha \beta}^{p_t} M_{\alpha \beta}(24)  \over t - m^2} 
  - {M_{\mu \nu}(14)N_{\mu \nu, \alpha \beta}^{p_u} M_{\alpha \beta}(23) \over u - m^2}   
  \nnl &+ &  C(1234) , 
\eea 
where $M_{\mu \nu}(jk)$ is defined by the decomposition of the 3-graviton amplitude:  
$\cM(jkl) \equiv  M_{\mu \nu}(jk) \epsilon_l^{\mu \nu}$. 
At this point the 4-point amplitude is determined up to contact terms, which will be constrained by requiring a specific high-energy behavior.

Let us comment on how \eref{DRGT_M3drgt} can be derived. 
The brute force way would be to take the cubic graviton terms in the dRGT Lagrangian and calculate the 3-point on-shell amplitude using the Feynman rules. 
A more intuitive way is the following. 
One can systematically build the 3-graviton amplitude as an expansion in the number of momentum insertions. 
At zero momentum insertion,  $\langle  \epsilon_1 \epsilon_2 \epsilon_3   \rangle$ is the unique Lorentz-invariant contraction of 3 polarization tensors. 
For scalar polarizations $\epsilon_i \sim E^2/m^2$ for $E \gg m$, 
thus the zero-momentum piece leads to the 4-point amplitude growing in the UV as $\cM(1^0 2^0 3^0 4^0) \sim E^6 \times E^6/E^2 = E^{10}$.  
For two momentum insertions there are two possible structures:  
$a_2 \langle  p_{23} \epsilon_1 p_{23} \rangle  \langle \epsilon_2 \epsilon_3 \rangle
+ b_2 \langle  p_{23} \epsilon_1 \epsilon_3 \epsilon_2 p_{13} \rangle$, together with their cyclic permutations. 
For generic $a_2$ and $b_2$, the 4-point amplitude will all scalar polarizations would grow as  $\cM(1^0 2^0 3^0 4^0) \sim E^8 \times E^8/E^2 = E^{14}$, much faster than that mediated by the zero-momentum-insertions term. 
However, a softer behavior is obtained if the $p/m$ terms in the numerator $N_{\mu \nu, \alpha \beta}$ annihilate $M(jk)$, and thus do not contribute to the amplitude.
This is equivalent to requiring that the two-momentum-insertion terms are invariant under the  transformation 
$\epsilon_j^{\mu\nu}\rightarrow \epsilon_j^{\mu\nu}\ + p_j^{\mu}\xi^\nu+\xi^\mu p_j^{\nu}$  for arbitrary $\xi$.
This fixes $b_2 = -2 a_2$. 
Finally, we set $a_2 = - 1/2 \mpl^2$ so as to recover the standard GR normalization in the massless limit. 
One could continue the EFT expansion of the 3-graviton amplitudes by adding terms with four and six momentum insertions. 
The former can be reduced to those with zero and two insertions by using momentum conservation and on-shell conditions~\cite{Bonifacio:2018vzv}. 
The latter would lead to amplitudes with transverse polarizations growing as $\cO(E^{10})$, and corresponds to deforming the dRGT gravity Lagrangian by a cubic term constructed out of the Weyl tensor~\cite{Ruhdorfer:2019qmk}. 
In this paper we restrict to the usual dRGT cubic graviton interactions described on-shell by \eref{DRGT_M3drgt}.

\subsection{UV behavior and contact terms}

We focus now on the high energy behavior of the four-graviton amplitude in \eref{DRGT_M4drgt}. 
As mentioned earlier, different graviton polarizations come with a different energy dependence for $E \gg m$:  
the $h=0$ component  is $\mathcal{O}(E^2)$, the $h = \pm 1$  components are $\mathcal{O}(E)$,  
while the transverse $h = \pm 2$ polarizations are $\mathcal{O}(1)$. 
Consequently, in the absence of the contact terms $C(1234)$ in  \eref{DRGT_M4drgt},  the worst possible UV behavior of different polarization amplitudes is estimated as 
\beq
\cM(0000) \sim \cO(E^{10}), \ \
\cM(1000) \sim \cO(E^{9}), \ \
\cM(1100) \sim \cO(E^{8}), \ \
\cM(2100), \cM(1110) \sim \cO(E^{7}), 
\eeq 
where we abbreviate $\cM(1^{h_1}2^{h_2}3^{h_3}4^{h_4}) \equiv \cM(h_1 h_2 h_3 h_4)$. 
The goal is to reduce the UV behavior down to  $\cO(E^{6})$ or better for all these amplitudes. 
To this end,  we introduce a basis of independent contact terms with zero and two momentum insertions:  
\bea
\label{eq:DRGT_contact}
C_1^{(0)} &= &  \langle \{ \epsilon_1,  \epsilon_2  \} \{ \epsilon_3 , \epsilon_4  \} \rangle
+ ({\rm x}),
\qquad C_2^{(0)}  =   \langle \epsilon_1  \epsilon_2 \rangle \langle \epsilon_3  \epsilon_4 \rangle
+ ({\rm x}),
\nnl 
C_{1}^{(2)} & = &    s \langle \{ \epsilon_1,  \epsilon_2  \} \{ \epsilon_3 , \epsilon_4  \} \rangle
+ ({\rm x}), \qquad 
C_{2}^{(2)}  =      \langle p_s \{ \epsilon_1,  \epsilon_2  \} \{ \epsilon_3 , \epsilon_4  \} p_s  \rangle
+ ({\rm x}),
\nnl 
C_{3}^{(2)} & = &   
   \langle  p_s \epsilon_1 \epsilon_3  \epsilon_2  \epsilon_4 p_s \rangle 
+ \langle  p_s \epsilon_2 \epsilon_3  \epsilon_1  \epsilon_4 p_s \rangle 
+ \langle p_s \epsilon_1 \epsilon_4  \epsilon_2  \epsilon_3  p_s\rangle  
+ \langle p_s \epsilon_2 \epsilon_4  \epsilon_1  \epsilon_3  p_s \rangle    + ({\rm x}),
\nnl 
C_{4}^{(2)} & = & s \left [  
\langle \epsilon_1  \epsilon_3 \rangle \langle \epsilon_2  \epsilon_4 \rangle
+ \langle \epsilon_1  \epsilon_4 \rangle \langle \epsilon_2  \epsilon_3 \rangle \right ]
+ ({\rm x}), \
C_{5}^{(2)}  =   
\langle p_s \epsilon_1 \epsilon_2 p_s \rangle \langle \epsilon_3 \epsilon_4 \rangle 
+  \langle  \epsilon_1 \epsilon_2  \rangle \langle p_s \epsilon_3 \epsilon_4 p_s \rangle   
+ ({\rm x}),
\nnl 
C_{6}^{(2)} & = &  
   \langle p_s \epsilon_1 \epsilon_3 p_s \rangle \langle \epsilon_2 \epsilon_4 \rangle 
+ \langle p_s \epsilon_2 \epsilon_3 p_s \rangle \langle \epsilon_1 \epsilon_4 \rangle 
+ \langle p_s \epsilon_1 \epsilon_4 p_s \rangle \langle \epsilon_2 \epsilon_3 \rangle 
+ \langle p_s \epsilon_2 \epsilon_4 p_s \rangle \langle \epsilon_1 \epsilon_3 \rangle   
+ ({\rm x}),
\eea 
where $({\rm x})$ stands for t- and u-channel crossed terms: $(2 \leftrightarrow 3) + (2 \leftrightarrow 4)$, 
and 
$\{ \eps_j, \eps_k \} \equiv (\epsilon_j^{\mu \nu} \epsilon_k^{\nu \rho} +  \epsilon_k^{\mu \nu} \epsilon_j^{\nu \rho})/2$.
There is no need to consider expressions with more than two momentum insertions, as they would lead to amplitudes growing faster than $\cO(E^{10})$. 
The contact terms can be parametrized as
\beq 
C(1234) = {1 \over  \mpl^2} \left [ m^2 \sum_{i=1}^2 w_i^{(0)} C_i^{(0)} + \sum_{i=1}^6 w_i^{(2)} C_i^{(2)} \right ],
\eeq 
and the Wilson coefficients $ w_i^{(n)}$ are chosen so as to reduce the UV behavior down to $\cO(E^6)$. 
This is achieved for the choice 
\bea
\label{eq:DRGT_wilson}
w_{1}^{(2)}   & =  & -4 , \qquad  w_{2}^{(2)}  =  8 , \qquad w_{3}^{(2)}  = -4, 
\qquad w_{4}^{(2)}  =    {a_0^2  - 1 \over 6}   , 
\qquad w_{5}^{(2)}  =  {a_0 (2 a_0  + 1) \over 6}  ,  
\nnl  w_{6}^{(2)}  &= & 2 ,
\qquad 
w_{2}^{(0)}  = {7 \over 2} 
-  {a_0 (2 a_0  + 1) \over 3}      - {1\over 2} d_0, 
\qquad w_{1}^{(0)} = d_0 . 
\eea 
This leaves two unconstrained parameters: $a_0$ from the 3-graviton amplitude \eref{DRGT_M3drgt}, and $d_0$ parametrizing a preferred direction in the space of  the contact terms in \eref{DRGT_contact}. 
They are related  to the commonly used parameters $c_3$ and $d_5$  in the dRGT Lagrangian~\cite{Hinterbichler:2011tt} via the map 
\beq
\label{eq:DRGT_drgtmap}
 a_0 = 3 (1 - 4 c_3) , \qquad 
 d_0 = (3 + 24 c_3 + 96 d_5). 
\eeq
The 4-graviton amplitude in \eref{DRGT_M4drgt} with the Wilson coefficients  adjusted as in \eref{DRGT_wilson} is the same as the one calculated directly (and more laboriously) from the dRGT Lagrangian.
In particular, the $\cM(0000)$, $\cM(1100)$, $\cM(1111)$, and $\cM(2000)$  amplitudes grow as  
$(E/\Lambda_3)^6$ in the UV, where 
\beq
\label{eq:DRGT_cutoff}
\Lambda_3  \equiv (m^2 \mpl)^{1/3} . 
\eeq
Here $\Lambda_3$ is the strong coupling scale where the graviton scattering amplitudes become non-perturbative. 
This also sets the highest possible cutoff scale of dRGT as long as no assumptions whatsoever are made about its UV completion.\footnote{%
If a Poincar\'{e} invariant, local, and causal UV completion is assumed, the maximum cutoff scale of dRGT  is $\Lambda_4 \equiv (m^3 \mpl)^{1/4}$ rather than $\Lambda_3$~\cite{Bellazzini:2017fep}, which is orders of magnitude smaller for realistic gravitino masses.
Extending the validity range of a massive gravity EFT all the way up to $\Lambda_3$ thus requires violation of established principles at macroscopic distance scale.  
We leave that option open in the following, without entering the discussion whether or not this is reasonable.  
}

\section{Matter coupling in dRGT}
\label{sec:matter}
\setcounter{equation}{0}

In this section we study interactions of the massive graviton with matter, that is with  particles of spin 0, 1/2, or 1.  
The strategy will be similar to the one employed for graviton self-interactions discussed in \sref{drgt}. 
We first write down the on-shell 3-point amplitudes 
$\cM(1 2 \mathbf{3})$ involving two matter particles and one graviton.
We focus on the amplitudes with the minimal number of momentum insertions, which are closely related to the minimal gravitational interactions of matter in GR.  
Then we construct the 4-point amplitude describing Compton scattering of matter on massive gravitons. 
Unitarity dictates that it must have the form 
\beq 
\label{eq:CS_ComptonGeneral}
\cM(1 \mathbf{2} 3 \mathbf{4}) =  
-    {\cM(1\hat p_s \mathbf{2}) \cM(3 p_s \mathbf{4}) \over s - M^2 }
 - {\cM(1 3 \mathbf{\hat p_t} ) \cM(\mathbf{2} \mathbf{4}\mathbf{p_t}) \over t - m^2} 
    - {\cM(1 \hat p_u \mathbf{4}) \cM(2 p_u \mathbf{3} ) \over u - M^2}   
    + C(1 \mathbf{2} 3 \mathbf{4}) , 
\eeq 
where the bold (unbold) entries correspond to the graviton of mass $m$ (matter particle of mass $M$), hats denote outgoing particles,   and eventual summation over polarizations of the intermediate particles $p_i$ is implicit.
The pole terms are schematically represented in \fref{Compton}.
Note that the t-channel depends also on the 3-graviton amplitude, which is assumed to be the one in \eref{DRGT_M3drgt}.  
Given the 3-point amplitudes, Compton scattering is determined up to the contact terms $C$. 
For the latter we assume the most general form in a systematic expansion in the number of momentum insertions. 
Finally, we study the Compton amplitudes for $E \gg m,M$.  
Massive gravity is an EFT, which is also reflected in Compton amplitudes growing for $m \ll E \ll \Lambda$, and eventually hitting strong coupling at a finite energy scale.
We discuss the possibility of adjusting the contact terms so as to soften the UV behavior and thus postpone the onset of strong coupling in the matter sector.  

\begin{figure}[tb]
\centering
\begin{minipage}{.3\textwidth}
  \centering
  \vspace{-0.5cm}
  \textbf{s-channel}
  
   \vspace{0.5cm}
  \includegraphics[width=.8\linewidth]{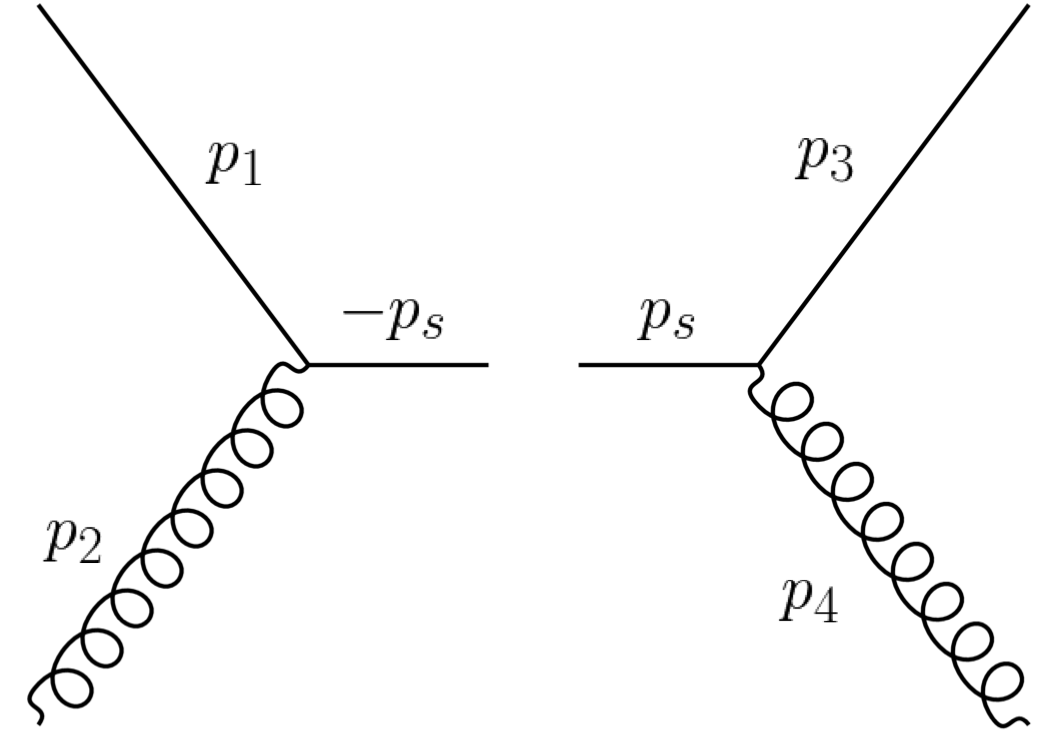}
\end{minipage}%
\begin{minipage}{.3\textwidth}
  \centering
  \vspace{-0.3cm}
  \textbf{t-channel}
  
   \vspace{0.3cm}

  \includegraphics[width=.5\linewidth]{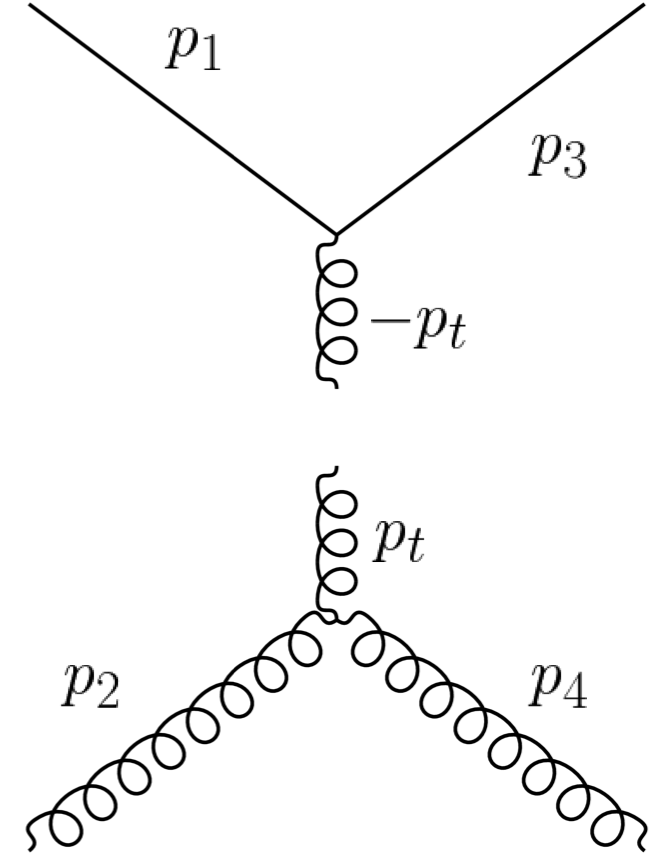}
\end{minipage}
\begin{minipage}{.3\textwidth}
  \centering
  \vspace{-0.3cm}
  \textbf{u-channel}
  
   \vspace{0.3cm}
  \includegraphics[width=.5\linewidth]{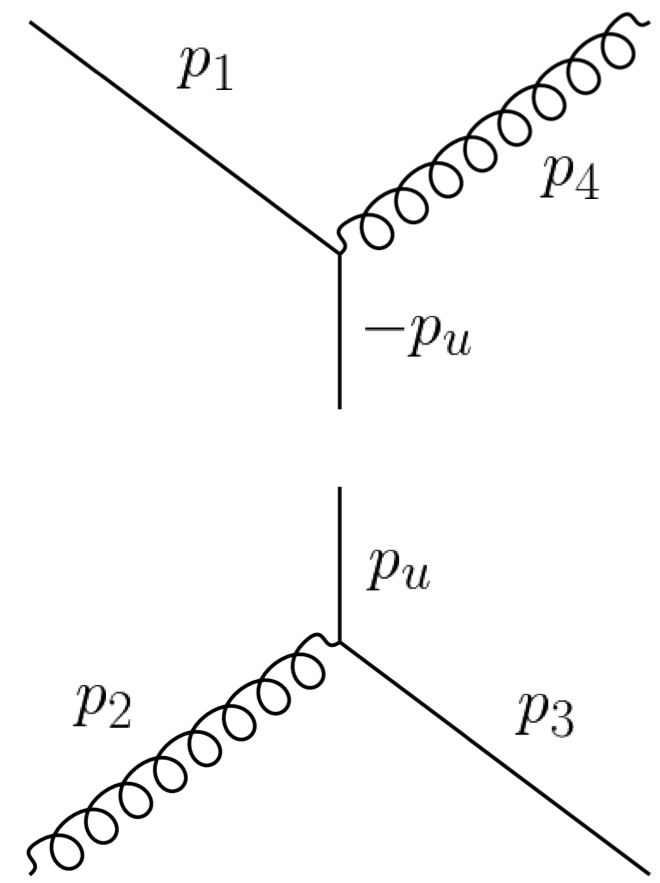}
\end{minipage}
\caption{\label{fig:Compton}
Schematic representation of the pole terms in the Compton amplitude in \eref{CS_ComptonGeneral}. }
\end{figure}

\subsection{3-point amplitudes}

We assume the following 3-point amplitudes for massless spin 0 scalars, spin 1/2 fermions,  and spin 1 photons interacting with gravity: 
\bea
\label{eq:CS_m3massless}
{\bf \rm Scalar:} & \quad  & \cM(1 2 {\mathbf 3}) =   - {c_s \over 2 \mpl}  \langle p_{12} \epsilon_3 p_{12} \rangle, 
\nnl 
{\bf \rm Fermion: }  & \quad  &  \cM(1^- 2^+ {\mathbf 3})  =    
  {c_f \over 2 \mpl } \langle j_{12} \epsilon_3 p_{12} \rangle ,
\qquad j_{12} \equiv (\lambda_1 \sigma^\mu \tilde \lambda_2), 
  \nnl
{\bf \rm Photon: }  & \quad  &  \cM(1^- 2^+ {\mathbf 3}) =   
- {c_\gamma\over 2 \mpl }
\langle j_{12} \epsilon_3 j_{12} \rangle
\eea  
where $\lambda_i$ and $\tilde \lambda_i$, $i =1,2$, are the helicity spinors associated with the massless four-momenta $p_i$, 
$\epsilon_3$ is the polarization tensor of the massive graviton, 
and $p_{ij} \equiv p_i - p_j$. 
For scalars, the above is the unique on-shell 3-point interaction with a spin-2 particle.
For fermions and photons we could also consider non-minimal  amplitudes where both matter particles have the same helicity, which however require more momentum insertions and are ignored in this discussion.  
The overall normalization $c_X$, $X= s,f,\gamma$, encodes the strength of gravitational interactions of the particle $X$. 
In GR, where the graviton is massless, internal consistency enforces the equivalence principle, that is $c_X =  1$ for any form of matter~\cite{Weinberg:1964ew}. 
In massive gravity $c_X$ are a priori free parameters, as there is no symmetry or unitarity arguments to fix them. 
In particular, there could be a distinct value of $c_X$ for different matter particles: electrons, quark, photons...    
 
For massive scalars, the 3-point amplitude remains exactly the same as in \eref{CS_m3massless}. 
On the other hand, for spin-1/2 fermions and spin-1 vectors it has to be modified to reflect the different little group transformation properties of massive particles.
For example for spin-1/2 it takes the form 
\beq
{\rm Massive} \ {\rm fermion: }  \quad   
\cM({\mathbf 1}_f  {\mathbf 2}_f \mathbf{3}_G)  =   {c_f \over 2 \mpl } (\chi_1 \sigma^\mu \tilde \chi_2 +  \chi_2 \sigma^\mu \tilde \chi_1) \epsilon_3^{\mu \nu} p_{12}^\nu .   
\eeq 
 In our analysis below we take matter to be massless, $M = 0$, and only briefly comment on what changes for $M>0$. 
 
\subsection{Compton scattering}
\label{sec:cs}

The next step is to calculate the Compton  amplitudes. 
The procedure is very similar for scalars, fermions, and photons.
Below we discuss the massless scalar case in some gory detail, while for fermions and photons we only present the final results. 

We are interested in the 4-point amplitude 
$\cM_c \equiv  \cM (1 {\mathbf 2}  3  {\mathbf 4})$, 
which we calculate using \eref{CS_m3massless} with  $M=0$. 
The residues of the pole terms are given by  
\bea
R_s \equiv  -\cM(1\hat p_s \mathbf{2}) \cM(3 p_s \mathbf{4})  & = &  -  {4 c_s^2 \over \mpl^2}  \langle p_1 \epsilon_2 p_1 \rangle  \langle p_3  \epsilon_4 p_3 \rangle , 
\nnl 
R_u \equiv - \cM(1 \hat p_u \mathbf{4}) \cM(2 p_u \mathbf{3} )   & = &  -  {4 c_s^2  \over \mpl^2}  \langle p_1 \epsilon_4 p_1 \rangle  \langle p_3  \epsilon_2 p_3 \rangle , 
\nnl 
R_t \equiv - \cM(1 3 {\mathbf{\hat p_t}} ) \cM(\mathbf{2} \mathbf{4}\mathbf{p_t}) & = &    {c_s \over 2 \mpl^2 }  p_{13}^\mu p_{13}^\nu
N_{\mu\nu,\alpha \beta}^{p_t} \mathcal{M}^{\alpha\beta}(24),  
\eea
where $N$ is given in \eref{DRGT_numerator},  
and  $M^{\alpha\beta}(24)$ is defined via the 3-graviton amplitude in \eref{DRGT_M3drgt}:
$\cM(\mathbf{i} \mathbf{j} \mathbf{k}) \equiv \epsilon_k^{\alpha \beta}  M_{\alpha\beta}(ij)$. 
Simple power counting shows that, for $E \gg m$,  the residues behave as $R_i \sim E^8$ when both gravitons have scalar polarizations. 
Their contributions to the scattering amplitude is  
$\cM_c \sim (E/\Lambda_3)^6$, $\Lambda_3 = (m^2 \mpl)^{1/3}$, which is the same high-energy behavior as for graviton self-scattering in dRGT.
Consequently,  Compton scattering becomes non-perturbative at the dRGT strong coupling scale, and a priori there is no need to fiddle with the  contact terms $C(1 \mathbf{2} 3 \mathbf{4})$  in \eref{CS_ComptonGeneral} so as to increase the  validity range of the EFT.  
Nevertheless, in the spirit of EFT we are interested in the completely general expression for the Compton amplitude, and for this reason we construct $C(1 \mathbf{2} 3 \mathbf{4})$ order by order in the EFT expansion.
We consider contact terms that do not worsen the UV properties, 
that is with up to two momentum insertions, contributing  $\cO(E^6)$ or softer to the amplitude. 
Up to this order, one basis of independent contact terms with correct little group transformations and Bose symmetry is  
\beq 
O^{(0)}_{1} =  m^2 \langle \epsilon_2 \epsilon_4 \rangle, \
O^{(2)}_{1} =   t \langle \epsilon_2 \epsilon_4 \rangle, \ 
O^{(2)}_{2} =  \langle  p_t \epsilon_2 \epsilon_4 p_t \rangle,   \ 
O^{(2)}_{3} =    \left ( \langle  p_s  \epsilon_2 \epsilon_4 p_s  \rangle
  +  \langle  p_u \epsilon_2 \epsilon_4 p_u \rangle \right ) . 
\eeq 
The general contact terms spanned by this basis,  
\beq
  C(1 \mathbf{2} 3 \mathbf{4}) = {1 \over \mpl^2} \left ( 
 c^{(0)}_{1} O^{(0)}_{1} + \sum_{k=1}^3 c^{(2)}_{k} O^{(2)}_{k} \right ),  
\eeq 
are included in the amplitude in \eref{CS_ComptonGeneral}. 

We now study the UV properties of the Compton amplitude  as a function of the dRGT parameter $a_0$, the scalar coupling strength $c_s$, and the four Wilson coefficients $c^{(n)}_{k}$.  
For a generic point in this parameter space, the Compton amplitude with both gravitons having scalar polarizations grows as $\cO(E^6)$ for $E \gg m$:
\beq 
\cM(1 \mathbf{2}^0 3 \mathbf{4}^0)= 
\frac{c_s(a_0 - 2) -12 c^{(2)}_{1} - 6 c^{(2)}_{2} + 2 c^{(2)}_{3} }{72 m^4 \mpl^2}t^3 
+ \frac{ 12 c_s^2 - 4 c_s  -4 c^{(2)}_{3} } {72 m^4 \mpl^2}(s^3+u^3) + \cO(E^4). 
\eeq 
It is clear that we can arrange the parameters so as to soften the UV behavior. 
For example, we can get rid of the $\cO(E^6)$ piece by fixing 2 Wilson coefficients as
$c^{(2)}_{2} = c_s^2  + {a_0 - 4 \over 6} c_s  - 2 c^{(2)}_{1}, 
\quad c^{(2)}_{3} = 3 c_s^2  - c_s$. 
In this restricted parameter space, the hardest Compton amplitude contains one scalar and one vector graviton polarization: 
\beq
\label{eq:CS_m410dgrt}
\cM(1 \mathbf{2}^{\pm 1} 3 \mathbf{4}^0) = 
\pm  { ( c_s  - c_s^2  ) \over 4\sqrt{3}}\sqrt{stu}(u-s) +  \cO(E^3). 
\eeq 
This amplitude cannot be softened by adjusting the Wilson coefficients, but it can be softened by fixing the coupling strength $c_s$ between the scalars and the graviton!  
Indeed, for $c_s =1$ the $\cO(E^5)$ piece vanishes.
This happens thanks to a cancellation between the s/u channels (which depend only on the scalar-graviton 3-point amplitude) and the t channel (which also depends on the 3-graviton amplitude). 
Note that $c_s=1$ is exactly the value  predicted by GR, where it is  required by virtue of the equivalence principle. 
In a way, massive gravity also discovers the equivalence principle, provided we require that the Compton scattering  is not harder than $\cO(E^4)$ in the UV.   
This is reminiscent of what happens in the theory of a  self-interacting massive spin-1 particle, where the Yang-Mills structure is discovered when we require that scattering amplitudes do not grow faster than  $\cO(E^2)$.  

Once  the $\cO(E^6)$ and $\cO(E^5)$ pieces are dealt with, the hardest amplitudes in the UV are the ones with two scalar or two vector polarizations: 
\begin{align}
    \begin{split}
\cM(1 \mathbf{2}^{0} 3 \mathbf{4}^0) 
&=-\frac{a_0 +2 c^{(0)}_{1} - 2 c^{(2)}_{1} }{12 m^2 \mpl^2}t^2+\frac{a_0 - 1}{12 m^2 \mpl^2}(s^2+u^2) + \cO(E^2)\\
\cM(1 \mathbf{2}^{1} 3 \mathbf{4}^1)  & =\frac{a_0 -6 c^{(2)}_{1}}{48 m^2 \mpl^2}t^2 + \cO(E^2).
    \end{split}{}
\end{align}{}
Those can be further softened by adjusting the Wilson coefficients $c^{(2)}_{2}$, $c^{(0)}_{1}$ {\em and} the free parameter $a_0$ in the 3-graviton amplitude in \eref{DRGT_M3drgt}. 
For the latter, the required value is $a_0 =1$ ($c_3 = 1/6$, in the standard conventions).
The complete set of parameters leading to the Compton amplitudes  
behaving as  $\cM_c \sim \cO(E^n)$ is 
\bea
\label{eq:CS_scalarsoft}
{E^5:} &\qquad &  
c^{(2)}_{2} = c_s^2  + {a_0 - 4 \over 6} c_s  - 2 c^{(2)}_{1}, 
\quad c^{(2)}_{3} = 3 c_s^2  - c_s;
\nnl 
{E^4:} &\qquad &  
c_s = 1 , \quad  c^{(2)}_{2} =  {a_0 + 2 \over 6}   - 2 c^{(2)}_{1}, 
\quad c^{(2)}_{3}  = 2;
\nnl 
{E^3:} &\qquad& 
c_s = 1 , \quad a_0 = 1, \quad 
c^{(0)}_{1} = -{1 \over 3}, \quad c^{(2)}_{1} = {1 \over 6}, \quad 
c^{(2)}_{2} = {1 \over 6}, \quad c^{(2)}_{3} = 2 . 
\eea  
At this point we have shot all the bullets. 
One can verify that for the parameters fixed as in the last line of  \eref{CS_scalarsoft}  one has 
$\cM(1 \mathbf{2}^{\pm 1} 3 \mathbf{4}^0) \sim \cO(E^3)$. 
Consequently, Compton scattering become non-perturbative at the scale  $\Lambda_c \sim (\mpl^2 m)^{1/3}$, which is far below the Planck scale, but well above the strong coupling scale $\Lambda_3$ of the pure graviton sector of dRGT.  
Compton amplitudes with other helicity configurations grow as $\cO(E^2)$ away from the forward limit,  
which is the same UV behavior as in GR.  

\vspace{1cm}

For Compton scattering of fermions or photons the story is the same. We skip the derivation and go directly to  the results. 
For two incoming fermions with opposite helicity, to calculate  $\cM(1^- \mathbf{2} 3^+ \mathbf{4})$ we use the 3-point amplitudes in \eref{CS_m3massless} and \eref{DRGT_M3drgt}, as well as the contact terms
$C(1^- \mathbf{2} 3^+ \mathbf{4}) = {1 \over \mpl^2} \sum_{k=1}^3 c^{(1)}_{k} O^{(1)}_{k}$ spanned by the basis 
\bea
\label{eq:CS_fcontact}
O^{(1)}_{1} & = &  (\lambda_1 \sigma^\mu \tilde \lambda_3) p_t^\nu \left ( \eps_2^{\mu \rho} \eps_4^{\nu \rho} +  \eps_4^{\mu \rho} \eps_2^{\nu \rho}  \right ) , 
\qquad 
O^{(1)}_{2} = 
(\lambda_1 \sigma^\mu \tilde \lambda_3) p_{13}^\nu \left (  \eps_2^{\mu \rho} \eps_4^{\nu \rho} +  \eps_4^{\mu \rho} \eps_2^{\nu \rho}  \right ) , 
 \nnl
O^{(1)}_{3} & = &   i \epsilon_{\mu\nu\alpha\beta} (\lambda_1 \sigma^\mu \tilde \lambda_3)  p_{24}^\nu
[\epsilon_2 \cdot \epsilon_4]^{\alpha \beta}. 
\eea 
The parameter space  consists of $c_f$, $a_0$ and the 3 Wilson coefficients $c^{(1)}_{k}$. 
Much as for scalars, for generic parameters the amplitude for scattering of fermions on the scalar graviton polarization grows like $\cO(E^6)$ in the UV. 
Although the number of Wilson coefficients is one smaller than in the scalar case, it remains possible to soften the Compton amplitudes all the way down to $\cO(E^3)$. 
The parameter settings leading to the growth not faster than $\cO(E^n)$ for $n=5,4,3$ are given by 
\bea 
E^5: \ && c^{(1)}_{2} =  {c_f - 2 c_f^2 \over 4} ;
\nnl 
E^4: \ && c_f  =1, \quad 
c^{(1)}_{1} = 0 , \quad 
c^{(1)}_{2} = -{1 \over 4}, \quad 
c^{(1)}_{3} = - {1 \over 4}; 
\nnl
E^3: \ && c_f = 1, \quad 
a_0 = 1, \quad 
c^{(1)}_{1} = 0 , \quad 
c^{(1)}_{2} = -{1 \over 4}, \quad 
c^{(1)}_{3} = - {1 \over 4}. 
\eea 
Again, demanding Compton amplitudes to be $\cO(E^4)$ or better leads to the equivalence principle, $c_f = 1$, while further softening of the UV behavior occurs for the special value of  the dRGT parameter $a_0 = 1$. 
This pattern is repeated for the amplitude with two incoming photons of opposite helicity and 2 massive gravitons. 
In this case there is a single contact term at the leading order:
\beq
\label{eq:CS_vcontact}
O_A = (\lambda_1 \sigma^\mu \tilde \lambda_3)  (\lambda_1 \sigma^\nu \tilde \lambda_3) 
\eps_2^{\mu \rho} \eps_4^{\nu \rho} , 
\eeq 
and the parameter space  consists of $c_\gamma$, $a_0$, and the Wilson coefficient $c_A$. The parameter settings leading to Compton amplitudes behaving as  $\cO(E^n)$ for $n=5,4,3$ are given by 
\bea 
\label{eq:CS_photonsoft}
E^5: \ && c_A = {c_\gamma(c_\gamma -1) \over 2 } ;
\nnl 
E^4: \ && c_\gamma  = 1, \quad  c_A  = 0 ; 
\nnl
E^3 : \ && c_\gamma = 1, \quad  a_0 = 1, \quad  c_A  = 0 . 
\eea
Once again $c_\gamma = 1$ and $a_0 = 1$ emerges as the special point where the Compton amplitudes are $\cO(E^3)$ or softer.  
For massless fermions and photons we also have same-helicity Compton amplitudes, e.g. $\cM(1^- \mathbf{2} 3^- \mathbf{4})$. 
In this case there is no pole contribution, given our assumption of minimal coupling in \eref{CS_m3massless}, 
however there can be a contribution from the contact term:  
${m \over \mpl^2}(\lambda_1 \lambda_3) \langle \epsilon_2 \epsilon_4 \rangle$ for fermions and ${1 \over \mpl^2}(\lambda_1 \lambda_3)^2 \langle \epsilon_2 \epsilon_4 \rangle$ for photons.
These lead to $\cM(1^- \mathbf{2}^0 3^- \mathbf{4}^0)$ growing as $\cO(E^5)$ ($\cO(E^6)$) for fermions (photons).   
The Wilson coefficients of these contact terms should be set to zero if we require $\cO(E^4)$ or better behavior of Compton amplitudes.

 \vspace{1cm}
 
The picture does not change if we consider matter particles with non-zero mass $M$.  
For massive spin 1/2 and spin 1 particles the contact terms have to be modified compared to \eref{CS_fcontact} and \eref{CS_vcontact} in order to reflect the correct little group transformation properties, 
and a larger set of contact terms needs to be considered.
Nevertheless, in all cases the qualitative features of Compton scattering on gravitons  do not differ from the massless case for $E \gg m,M$. 
It is of course intuitively expected that the UV properties of scattering amplitudes are insensitive to the masses of matter particles.

\subsection{Discussion}
\label{sec:disc}

The pure gravity sector of dRGT depends on the graviton mass $m$ and two free parameters $a_0$ and $d_0$ ($c_3$ and $d_5$ in the standard conventions).
Once matter is taken into account, the parameter space is much enlarged. 
It includes the coefficients of the 3-point amplitudes describing the graviton coupling to matter, and those of the 4-point contact terms between matter and gravitons.
For the minimal coupling in \eref{CS_m3massless}, the 3-point amplitude $\cM(XXG)$ for each matter particle $X$ is characterized by a single parameter $c_X$, which can be interpreted as the relative coupling strength compared to that of the massless graviton in GR. 
In \sref{cs} we studied the  Compton scattering amplitudes $\cM_c = \cM(X G \to X G)$ in massive dRGT gravity minimally coupled to matter as a function of $c_X$ and the 4-point contact terms. 
A number of interesting properties  was uncovered: 
\bi
\item For a generic point in the parameter space, the amplitudes grow with energy as $\cM_c \sim (E/\Lambda_3)^6$, 
where $\Lambda_3$ is the strong coupling scale of pure dRGT gravity defined in \eref{DRGT_cutoff}. 
Thus, $\cM_c$ are always perturbative below the dRGT cutoff, and a priori no adjustment of the parameters is needed to soften their UV behavior.  
\item 
Nevertheless, the UV behavior of the Compton amplitudes can be considerably softer in some regions of the parameter space. 
For generic $a_0$ in the pure gravity sector, one can achieve 
$\cM_c \sim E^4/m^2 \mpl^2$. 
In such a case, the Compton amplitudes hit the strong coupling at the scale $\tilde \Lambda_c$ defined as
\beq
\tilde \Lambda_c = \sqrt{m \mpl}, 
\eeq
which is many orders of magnitude larger than the dRGT strong coupling scale $\Lambda_3$ for graviton masses of phenomenological interest.
The softer behavior is possible thanks to cancellations between $s/u$- and $t$-channel diagrams.
(As an aside note,  the $t$-channel exists thanks to the 3-graviton amplitude, thus the cancellation would not be possible for a spin-2 theory without the cubic self-interaction.) 
To arrive at $\cM_c \sim E^4$, one needs to adjust parameters in the matter sector.
Apart from fixing the contact terms, also the coupling strength $c_X$ between the massive graviton and matter has to be set to the GR value $c_X=1$.
Recall that in GR $c_X=1$ is required 
by the absence of unphysical poles in tree-level Compton amplitudes~\cite{Benincasa:2007xk,McGady:2013sga}.
In massive gravity there is no such consistency condition, and thus any value of $c_X$ is allowed from the EFT point of view. 
This is at odds with experimental facts that firmly establish the equivalence principle~\cite{Touboul:2017grn}, that is $c_X \approx 1$ for all types of matter to a fantastic accuracy. 
It is intriguing that the equivalence principle emerges in massive gravity as well, simply by demanding a softer UV behavior of $\cM_c$. 
\item 
This is not all.
We found that that the Compton amplitudes in massive gravity can be further softened by adjusting one more parameter, namely $a_0$ parametrizing the 3-graviton amplitude in \eref{DRGT_M3drgt}. 
After setting $a_0 = 1$ (and eventually adjusting some other Wilson coefficients),  the UV behavior is softened by another notch to $\cM_c \sim (E/\Lambda_c)^3$. 
The new scale is defined as 
\beq
\Lambda_c = (m \mpl^2)^{1/3}. 
\eeq 
For viable graviton masses there is a strong hierarchy $\Lambda_c \gg \tilde \Lambda_c \gg \Lambda_3$. 
For example, for $m = 10^{-32}$~eV, we have 
$\Lambda_3 \approx (300~{\rm km})^{-1}$, 
$\tilde \Lambda_c \approx (0.04~{\rm mm})^{-1}$, 
$\Lambda_c \approx (5 \times 10^{-12}\, {\rm mm})^{-1}$.  
Note that $(a_0,d_0) = (1,5)$ (or $(c_3,d_5) = (1/6,-1/48)$) is a special point in dRGT, leading to non-interacting scalar polarizations in the decoupling limit ($\mpl \to \infty$ with $\Lambda_3$ held fixed) of dRGT~\cite{deRham:2010ik}. 
This specialness has little consequence in the pure gravity sector: while $\cM(0000)$ and $\cM(2000)$ self-scattering amplitudes are softened for this parameter choice, 
since $\cM(1111)$ and $\cM(1100)$ graviton  still grow as $(E/\Lambda_3)^6$.  
On the other hand, the specialness of $a_0 = 1$ is well visible in the matter sector, allowing for maximally soft gravitational Compton scattering amplitudes. 
\ei
The features described above are universal for all types of matter: spin-0 scalars, spin-1/2 fermions, spin-1 vectors, and independent of whether these particles are massive or massless. 

What is the significance of these findings? 
At face value, they do not change the fate of dRGT gravity. 
Even though $\cM_c \sim (E/\Lambda_c)^3$ at tree level, the cutoff $\Lambda$ of the full theory is much lower than $\Lambda_c$, namely  $\Lambda \lesssim \Lambda_3$ or even $\Lambda \lesssim  \Lambda_4 \ll \Lambda_3$ if a Poincar\'{e} invariant, local, and causal UV completion is assumed.
Above $\Lambda$ there are 2 options: 
either new weakly coupled degrees of freedom are introduced, 
or the graviton polarizations $0$ and $\pm 1$   become strongly coupled.  
In either case  the EFT cannot be used in its current form. 

Furthermore, the lower scale $\Lambda_3$ may feed into $\cM_c$ at the loop level, and in this paper we have not shown that this can be tamed by fixing EFT parameters. 
On a related note, the $2 \to n$ amplitudes $\cM(XX \to G \dots G)$ with $n > 2$ gravitons may hit the strong coupling faster than at $\Lambda_c$, unless again their behavior can be tamed by $n+2$-point contact terms.  

Nevertheless, we find it intriguing that the softness of the  Compton amplitudes is intimately connected to the equivalence principle. 
As the latter is indispensable for any phenomenological applications, it is tempting to think that the former should be  an essential ingredient of massive gravity. 
What this softness may buy us depends on the scenario. 
If $\Lambda \ll \Lambda_3$, then the theory is completed with new weakly coupled degrees of freedom with masses of order $\Lambda$. 
The fact that matter scattering amplitudes need not be regulated at these scales implies that the new degrees of freedom need not be coupled to matter, which may help in construction of phenomenologically viable theories.

If, on the other hand, $\Lambda \sim \Lambda_3$, the possible  advantage is that matter scattering amplitudes are still deep in the perturbative regime as the theory approaches the dRGT cutoff. 
One can speculate that, at $\Lambda_3$, the pure gravity sector undergoes a phase transition which suppresses propagation of the scalar and vector polarizations. Above that scale we deal with a theory of matter and transverse graviton polarizations {\em weakly coupled} to a strongly interacting sector comprised of $|h| < 2$ polarizations (and possibly other degrees of freedom).
This resembles the more familiar example of processes with  the SM particles at energies below the QCD confinement scale. 
While not all observables can be calculated from first principles in such theories, their perturbative expansion in weak couplings is still under control, and the uncertainties due to the presence of the  strong sector can be quantified.

\section{Positivity}
\label{sec:pos}

Additional constraints on the parameter space of an EFT can be obtained provided its UV completion is local, causal, and respects Poincar\'{e} invariance~\cite{Adams:2006sv}. 
Under these assumptions, a sum of certain low-energy residues of a forward, crossing-symmetric amplitude has to be strictly positive, which leads to inequalities that need to be satisfied by the EFT parameters.\footnote{See also \cite{deRham:2017avq,deRham:2017zjm} for positivity bounds on amplitudes beyond the forward limit.}  
In this section we first review the positivity bounds on the dRGT parameters in the pure gravity sector~\cite{Cheung:2016yqr}. 
Then we derive novel positivity bounds on the parameters characterizing the interactions between dRGT gravity and matter.   

\subsection{Pure gravity sector}
\label{sec:POS_pure}

Positivity places non-trivial constraints on the parameters $a_0$, $d_0$ of dRGT gravity~\cite{Cheung:2016yqr}. 
Consider the forward limit of the 2-to-2 graviton self-scattering amplitude:  
$\cM_F^{G_1 G_2}(s) \equiv \cM (G_1 G_2 \to G_1 G_2)|_{t=0}$.   
A crossing-symmetric $\cM_F$ calculated in dRGT behaves as
$\cM_F^{G_1 G_2}(s) = \Sigma^{G_1 G_2} s^2 + \cO(s^0)$ for $s \gg m^2$. 
The existence of a local, causal, and Poincar\'{e} invariant UV completion  then implies that $\Sigma^{G_1 G_2}$  must be strictly positive~\cite{Adams:2006sv,Bellazzini:2015cra,Cheung:2016yqr}:
\beq
\label{eq:POS_gravcond}
\Sigma^{G_1 G_2} > 0.  
\eeq
This holds for any states $G_i$, whether helicity eigenstates or combinations thereof, as long as the forward amplitude is crossing-symmetric, that is 
$\cM_F^{G_1 G_2}(4m^2- s) = \cM_F^{G_1 G_2}(s)$. 

We take the polarization vectors describing $G_i$ to be general  combinations of {\em linear} polarization eigenstates $\epsilon^h(p1)$:   
\beq
\epsilon_1 = \sum_h \alpha_h \epsilon^h(p_1), 
\qquad 
\epsilon_2 = \sum_h \beta_h \epsilon^h(p_2) ,
\eeq 
where $h \in (S,V,V',T,T')$. 
Working with linear polarizations is convenient because crossing symmetry is most transparent in this basis~\cite{Bellazzini:2015cra}. 
The coefficients $\alpha_h$, $\beta_h$ can be complex, while  the 4-vectors  $\epsilon^h(p_i)$ are all real. 
The explicit form of $\epsilon^h(p_i)$ is given e.g. in Ref.~\cite{Bellazzini:2017fep}. 
For our discussion the important point is that  
\beq 
\label{eq:POS_epscrossing}
\epsilon^h(p) = (-1)^h \epsilon^h(-p),
\eeq 
where we {\em define} $(-1)^h \equiv 1$ for $h =S,T,T'$,
and $ (-1)^h \equiv - 1$  for $h =V,V'$.  
Notice that $\cM_F$ is not crossing-symmetric for arbitrary $\alpha_i$, $\beta_i$. 
Indeed, for {\em all incoming} linear polarizations Bose symmetry under interchanging $1 \leftrightarrow 3$ gravitons requires  $\cM^{h_1 h_2 h_3 h_4}(s,t,u) = \cM^{h_3 h_2 h_1 h_4}(u,t,s)$. 
Then for 2-to-2 scattering of linear polarizations \eref{POS_epscrossing} implies that 
\beq 
\cM^{h_1 h_2 \to h_3 h_4}(s,t,u) = (-1)^{h_1+h_3}\cM^{h_3 h_2 \to h_1 h_4}(u,t,s).
\eeq 
It follows that the forward amplitude is automatically crossing symmetric for scattering of definite linear polarizations: 
$\cM^{h_1 h_2 \to h_1 h_2}(s)  = \cM^{h_1 h_2\to h_1 h_2}(u) = \cM^{h_1 h_2\to h_1 h_2}(4 m^2 - s)$. 
However that is not true for scattering of a general combination of linear polarizations: 
\beq
\cM_F^{G_1 G_2}(s) =  \sum_{h_1 \dots h_4} 
\alpha_{h_1} \alpha_{h_3}^* \beta_{h_2} \beta_{h_4}^*  
\cM^{h_1 h_2 \to h_3 h_4}(s) 
= \sum_{h_1 \dots h_4} (-1)^{h_1} (-1)^{h_3}
\alpha_{h_1}^* \alpha_{h_3} \beta_{h_2} \beta_{h_4}^* 
\cM^{h_1 h_2 \to h_3 h_4}(4m^2-s) . 
\eeq
It follows that $\cM_F^{G_1 G_2}(s)$ is crossing symmetric 
if $\alpha_h$ is real for $h=S,T,T'$ and 
$\alpha_h$ is purely imaginary for $h=V,V'$. 
The analogous condition holds for $\beta_h$. 
We thus calculate $\Sigma^{G_1 G_2} (a_0,d_0,\alpha_h,\beta_h)$ 
and minimize it  over $\alpha_h$, $\beta_h$ subject to these conditions.
The $(a_0,d_0)$ pairs for which the minimum is negative or zero are excluded.

\begin{figure}[tb]
\bc
  \includegraphics[width=0.7 \textwidth]{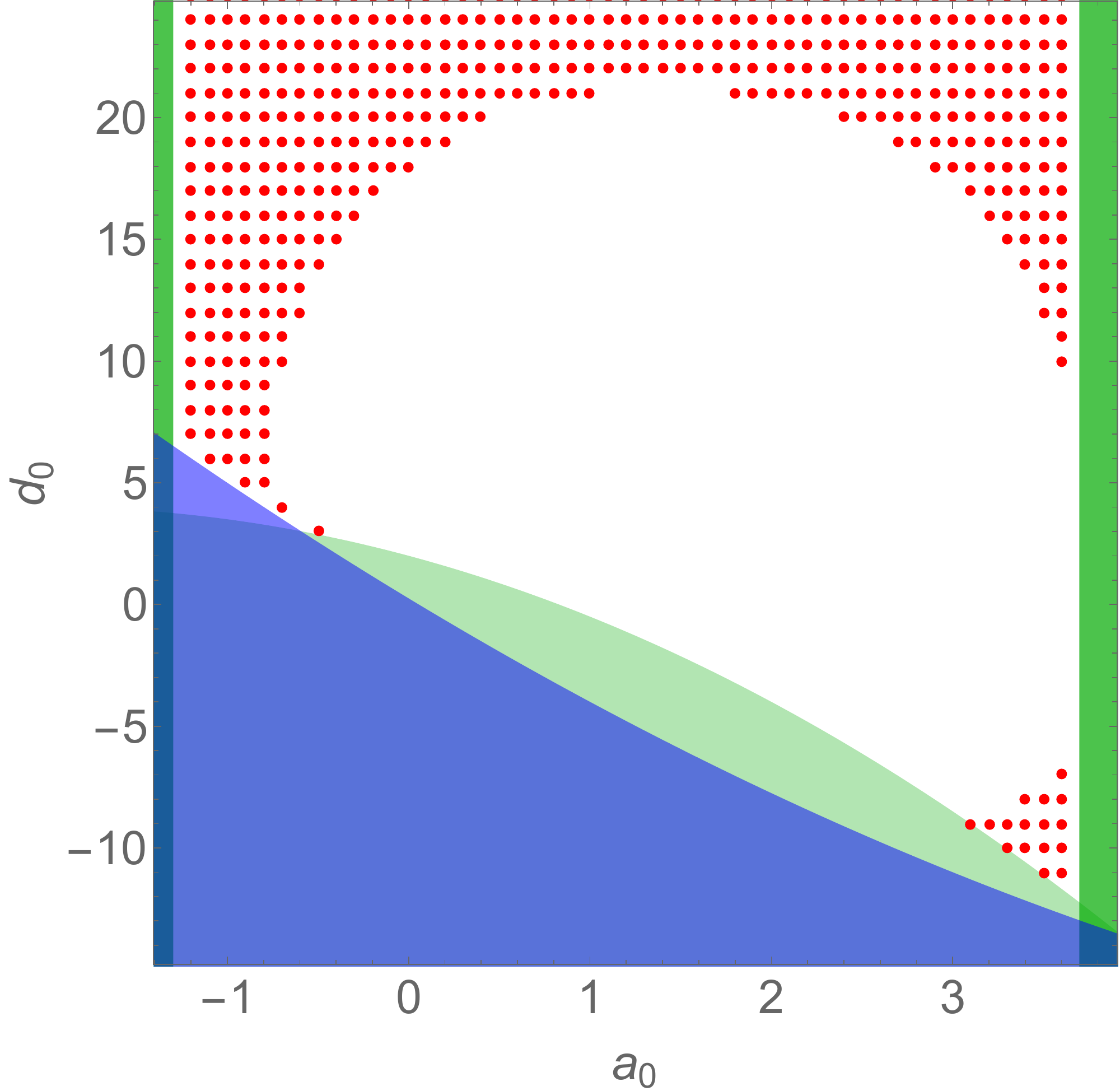}
  \caption{
    \label{fig:POS_drgt}
Positivity bounds on the dRGT parameter space.
Solid regions are excluded by forward scattering of definite linear polarizations of graviton: $VV$ (darker green), $VV'$ (lighter green), and $SS$ (blue). 
Scattering of other definite linear polarization states leads to weaker bounds, not shown in this figure.  
The red-dotted region shows additional constraints from scattering of general combinations of linear polarizations. 
  }
  \ec 
\end{figure}

The resulting constraints on the dRGT parameter space are shown in \fref{POS_drgt}. 
They agree with Ref.~\cite{Cheung:2016yqr} up to the change of variables in \eref{DRGT_drgtmap}.
An island of parameters remains allowed by these standard  positivity bounds.  
In particular, the parameter $a_0$ characterizing the 3-graviton amplitude, cf. \eref{DRGT_M3drgt}, is limited to the interval 
\beq
\label{eq:POS_a0range}
-0.7 \lesssim a_0 \lesssim 3.7, 
\eeq 
(or $-0.06 \lesssim c_3 \lesssim 0.31$).
The island is erased by the beyond-positivity bounds of Ref.~\cite{Bellazzini:2017fep}, unless the graviton self-scattering amplitudes calculated in this EFT are not valid above the scale $\Lambda$ satisfying the inequality 
\beq
\Lambda \lesssim \Lambda_4 \equiv (m^3 \mpl)^{1/4}. 
\eeq 
For realistic graviton masses $\Lambda_4$ is many orders of magnitude lower than $\Lambda_3$ in \eref{DRGT_cutoff}.  
For example, for $m = 10^{-32}$~eV, 
$\Lambda_4^{-1} \approx 3 \times 10^{7}$~km is an astronomical distance scale. 
We discuss beyond-positivity bounds in more detail later on, when discussing matter-gravity couplings.

\subsection{Matter-gravity couplings}

Positivity also constrains the parameters describing the interactions of the massive graviton with matter. 
This time the relevant object is the Compton scattering amplitude  in the forward limit:
$\cM_F^{XG}(s) \equiv  \cM(X G \to X G)|_{t=0}$, where $X$ stands for scalar, fermion, or vector matter particles. 
For a crossing-symmetric $\cM_F^{XG}(s)$, 
the low-energy residues should satisfy 
\beq
\label{eq:POS_matcond0}
{\rm Res}_{s\to M^2} {\cM_F^{XG}(s) \over (s - \mu^2)^3} 
+{\rm Res}_{s\to 2m^2 + M^2} {\cM_F^{XG}(s) \over (s - \mu^2)^3}
+ {\rm Res}_{s\to \mu^2} {\cM_F^{XG}(s) \over (s - \mu^2)^3} > 0,  
\eeq 
for any $\mu^2 \in [(m-M)^2,(m+M)^2]$. 
If $\cM_F^{XG}(s)$ calculated in the EFT does not grow faster than $\cO(s^2)$ in the UV, as is the case here, then the sum 
in \eref{POS_matcond0} can be traded for the residue in infinity. 
Expanding $\cM_F^{XG}(s) = \Sigma^{XG} s^2 + \cO(s^0)$, 
the coefficient $\Sigma^{XG}$ has to be strictly positive: 
\beq
\label{eq:POS_matcond}
\Sigma^{XG} > 0. 
\eeq 
This condition has the same form as for graviton self-scattering, cf. \eref{POS_gravcond}, in particular it is independent of the auxiliary parameter $\mu$.   

Below we consider matter scattering on a general combination of linear graviton polarizations: 
$\epsilon_2 = \sum_h \alpha_h \epsilon^h(p_2)$, $h \in (S,V,V',T,T')$.
Unlike for graviton self-scattering discussed in \sref{POS_pure}, for Compton scattering the amplitude is crossing-symmetric for an arbitrary choice of $\alpha_h$: 
$\cM_F^{XG}(s) = \cM_F^{XG}(2 M^2 + 2 m^2 - s)$.
We start with the positivity bounds on the massive graviton couplings to scalars, $X = S$.
The general EFT parameter space, where non-forward Compton amplitudes grow as $\cO(E^6)$ in the UV, consists of the graviton-scalar coupling strength $c_s$, and the Wilson coefficients of the graviton-scalar contact terms $c^{(0)}_{1}$, $c^{(2)}_{i}$, $i=1\dots 3$.  
Moreover, the Compton amplitudes depend also on the parameter $a_0$ characterizing the 3-graviton amplitude in dRGT.   
We find 
\beq
\label{eq:DRGTs_Mscf}
\Sigma^{SG}  =  {1 \over m^2 \mpl^2} \bigg \{ 
 c_s \left (|\alpha_T|^2 + |\alpha_{T'}|^2 \right )
+ {c_s (a_0+ 4) - c^{(2)}_{3} \over 4}  \left (|\alpha_V|^2 + |\alpha_{V'}|^2 \right )
+ {c_s^2 +  (a_0 + 3) c_s    -  c^{(2)}_{3} \over 3} |\alpha_S|^2
\bigg \}  . 
\eeq
Note that the high-energy limit of $\cM_F^{SG}$ is independent of the scalar mass $M$. 
The positivity bounds deduced from \eref{DRGTs_Mscf} are 
\beq
\label{eq:DRGTs_posE6}
{E^6:} \qquad c_s > 0, \qquad  c^{(2)}_{3} <  c_s (a_0+ 4)  , \qquad    c^{(2)}_{3} < c_s^2 +  (a_0 + 3) c_s   . 
\eeq 
We obtain a sharp result for the scalar coupling to massive gravitons: the overall coefficient $c_s$ of the 3-point amplitude in \eref{CS_m3massless} has to be strictly positive. 
Fortunately, the GR value $c_s=1$ is consistent with positivity. 
Furthermore, positivity of graviton self-scattering implies $a_0 + 4 >0$, from which it follows that the Wilson coefficient $c^{(2)}_{3}$ has to be strictly negative:  
$c^{(2)}_{3}< 0$. 
Other Wilson coefficients in the scalar-graviton sector are not subject to positivity bounds. 
This is because $O^{(2)}_{1}$, $O^{(2)}_{2}$, and the  $\cO(s^2)$ contribution of $O^{(0)}_{1}$ vanish in the forward limit. 

Positivity bounds become simpler in the parameter region where the Compton amplitudes are softer for $m \ll E \ll \Lambda$. 
Softening the amplitude down to $\cO(E^5)$ requires setting $c^{(2)}_{3} = 3 c_s^2  - c_s$, cf. \eref{CS_scalarsoft}. 
Then \eref{DRGTs_posE6} reduces to: 
\beq
\label{eq:DRGTs_posE5}
{E^5:} \qquad c_s > 0, \qquad   a_0 > 3 c_s - 5  , \qquad a_0 > 2 c_s - 4 . 
\eeq
Given the upper bound  $a_0 \lesssim 3.7$ in \eref{POS_a0range},   \eref{DRGTs_posE5} yields an upper bound $c_s \lesssim 2.9$.

To go further, softening Compton amplitudes down to $\cO(E^4)$ is possible only for $c_s = 1$. 
Then the first inequality in \eref{DRGTs_posE5} is moot, 
while the remaining two reduce to a single one 
\beq
\label{eq:DRGTs_posE4}
{E^4:} \qquad a_0 > - 2. 
\eeq
This is in fact weaker than the positivity bound on $a_0$ in \eref{POS_a0range} arising from graviton self-scattering. 
Softening the Compton amplitudes even further requires setting $a_0 = 1$, in which case \eref{DRGTs_posE4} is moot. 
Thus, in the parameter region where the Compton amplitudes behave as $\cO(E^3)$ for $m \ll E \ll \Lambda$,  positivity of the forward Compton amplitudes is automatically fulfilled.

For spin 1/2 and 1 matter particle the derivation of the positivity bounds is analogous. 
We only quote the final results for different levels of the EFT where the Compton amplitude behaves as $\cO(E^n)$, $n=6,5,4,3$. For spin-1/2 fermions we find  
\bea
\label{eq:DRGTf_pos}
{E^6:} &\qquad & c_f > 0, \qquad   
-  c_f^2  + (a_0 + 4) c_f   +  4 c^{(1)}_{2}  > 0 , \qquad 
(a_0 + 3) c_f + 4 c^{(1)}_{2}   > 0,
\nnl 
{E^5:} &\qquad & c_f > 0, \qquad   a_0 > 3 c_f - 5  , \qquad a_0 > 2 c_f - 4 , 
\nnl 
{E^4:} &\qquad & a_0 > - 2,
\eea 
while for spin-1 matter we find  
\bea 
\label{eq:DRGTv_pos}
{E^6:} &\qquad & c_\gamma > 0, \qquad
- 2 c_\gamma^2 +  (a_0 + 4) c_\gamma   > 2 c_A, \qquad
-  c_\gamma^2  + (a_0 + 3) c_\gamma    > 2 c_A, 
\nnl 
{E^5:} &\qquad & c_\gamma > 0, \qquad   a_0 > 3 c_\gamma - 5  , \qquad a_0 > 2 c_\gamma - 4 , 
\nnl 
{E^4:} &\qquad & a_0 > - 2. 
\eea 
In all cases, positivity is automatically fulfilled in the parameter region where Compton amplitudes grow as $\cO(E^3)$ for $m \ll E \ll \Lambda$. 

One thing that is striking about the positivity bounds is that they are universal for all matter particles, irrespectively of their mass and spin.  
In all cases they fix the sign of the gravity-matter coupling strength $c_X$, and they become moot when the Compton amplitudes are softened to $\cO(E^4)$. 
Furthermore, in our basis, they are sensitive to only a single contact term. 
Finally, for some values of $c_X$ and the relevant contact terms, they may imply new constraints on the dRGT parameter $a_0$, in addition to those imposed by forward graviton self-scattering discussed in \sref{POS_pure}.

We also comment on the positivity constraints on the amplitude for matter scattering $\cM(X X \to X X)$. 
For simplicity we assume $X$ does not have electric charge, so that the forward amplitude 
$\cM_F^{XX}(s) \equiv  \cM(X X \to X X)|_{t=0}$ 
is well defined. 
The massive graviton exchange results in the poles of $\cM_F^{XX}(s)$ at $s = m^2$ (s-channel) and $s = 4M^2 - m^2$ (u-channel).  
Positivity then requires 
\beq
\label{eq:POS_matter}
{\rm Res}_{s\to m^2} {\cM_F^{XX}(s) \over (s - \mu^2)^3} 
+{\rm Res}_{s\to 4 M^2 - m^2} {\cM_F^{XX}(s) \over (s - \mu^2)^3}
+ {\rm Res}_{s\to \mu^2} {\cM_F^{XX}(s) \over (s - \mu^2)^3} > 0  
\eeq 
for any $\mu^2 \in [0,4 M^2]$, 
which is equivalent  to $\Sigma^{XX} > 0$ if 
$\cM_F^{XX}(s) = \Sigma^{XX} s^2 + \cO(s^0)$. 
In the case at hand by direct calculation one finds 
$\Sigma^{XX} = {c_X^2 \over m^2 \mpl^2}$, 
therefore positivity of $\cM_F^{XX}(s)$ is trivially satisfied for any matter-graviton coupling.

\subsection{Beyond positivity}

Refs.~\cite{Bellazzini:2016xrt,Bellazzini:2017fep} observed that the coefficient of the $s^2$ term in the  forward amplitude should be not only positive, but also larger than a certain integral of the total cross section. 
The latter can be a large number when non-forward amplitudes grow fast for $m \ll E \ll \Lambda$, in which case positivity bounds can be substantially strengthened.
In the case at hand, 
the coefficient $\Sigma^{XG}$ of the $s^2$ term in the UV expansion of the forward Compton scattering amplitude  $\cM(X G \to X G)$ calculated in the EFT  must  satisfy
\beq
\label{eq:POS_bp}
\Sigma^{XG} > {1 \over 2 \pi} \int_{(m+M)^2}^\infty ds   \left [
{1 \over (s- \mu^2)^3} + {1 \over (s + \mu^2 - 2 m^2 - 2M^2)^3}
\right ] \sum_f {\int d\Pi_n} |\cM(X G \to f_n)|^2 , 
\eeq 
where the sum is over all possible $n$-body final states, and $d \Pi_n$ denotes the n-body phase space element. 
Above, $\cM(X G \to f_n)$ are the elastic and inelastic amplitudes in the full theory, which however can be approximated by the corresponding EFT expressions for $\sqrt{s}$ below the EFT cut-off $\Lambda$. 
If the right-hand side is large, the condition
on $\Sigma^{XG}$ is much stronger than mere positivity. 
This is referred to as {\em beyond-positivity} constraints in Ref.~\cite{Bellazzini:2017fep}. 

In the following we consider Compton scattering on a definite linear polarization state of the graviton. 
We focus here on scattering of massless photons, just  because the number of parameters is the smallest in this case and the formulas are concise; however the discussion is similar for scalars and fermions. 
The relevant parameter space consists of the graviton mass $m$, the dRGT parameter $a_0$,  
the photon-gravity coupling strength $c_\gamma$, and the Wilson coefficient $c_A$ of the leading 2-graviton-2-photon contact term.  
For scattering on the scalar polarization, the left-hand side of \eref{POS_bp} is 
\beq
\Sigma^{\gamma G^0} = 
{1 \over 3 m^2 \mpl^2} \left [ c_\gamma(3 + a_0 - c_\gamma) - 2 c_A \right ]. 
\eeq 
For the right-hand side we restrict to 2-body final states. 
Then the leading low-energy contribution to the integral comes from $\cM(\gamma G^0 \to \gamma G^0)$, which grows as $s^3$ for $m \ll \sqrt{s} \lesssim \Lambda$.
We can thus estimate the upper bound on the right-hand side: 
\beq
{\rm r.h.s} < 
{ (2 c_A + c_\gamma- c_\gamma^2)^2 \Lambda^8
\over 34560 \pi^2 m^8 \mpl^4 }. 
\eeq 
The beyond-positivity bound thus read 
\beq
 c_\gamma(3 + a_0 - c_\gamma) - 2 c_A  > 
{(2 c_A + c_\gamma- c_\gamma^2)^2 
\over 11520  \pi^2 }  {\Lambda^8 \over \Lambda_4^8},  
\eeq 
where $\Lambda_4 = (m^3 \mpl)^{1/4}$. 
This condition can be satisfied in only two ways. 
For generic $c_A$ and $c_\gamma$, 
we need $\Lambda \lesssim \Lambda_4$. 
This corresponds to $\Lambda^{-1}$ being an astronomical distance scale, which restricts the usefulness of this EFT as a theory of gravity. 
The other way is to set 
$2 c_A + c_\gamma- c_\gamma^2=0$. 
This is of course exactly the first condition in \eref{CS_photonsoft} required to soften $\cM(\gamma G \to \gamma G)$ from $\cO(E^6)$ down to $\cO(E^5)$. 
Thus, the beyond-positivity bounds provide another rationale for restricting the EFT parameter space, 
so as to arrive at softer Compton amplitudes! 
This softening is necessary if our  matter-gravity interactions are to emerge from a local, causal, and Poincar\'{e} invariant UV completion above a reasonably high cutoff scale. 

Similarly, the beyond-positivity bound on $\Sigma^{\gamma G^{\pm 1}}$ can be satisfied either for $\Lambda \lesssim \Lambda_3 = (m^2 \mpl)^{1/3}$,  or by setting  $c_\gamma = 1$ so as to avoid a large contribution of $|\cM(\gamma G^{\pm 1} \to \gamma G^0)|^2$ on the right-hand side of \eref{POS_bp}.
Once the Compton amplitudes are softened down to $\cO(E^4)$, the beyond-positivity bounds become equivalent in practice to the standard positivity bounds.  
The final comment is that, once beyond-positivity bounds on $\cM(X G \to X G)$ are satisfied, 
those on $\cM(X X \to X X)$ are automatically satisfied too.

\section{Conclusions}
\label{sec:con}
\setcounter{equation}{0}

In this paper we discussed interactions of matter particles with massive gravitons using the on-shell amplitude framework. 
We assume that graviton self-interactions are described by the dRGT gravity. 
That theory is characterized by the graviton mass $m$ and two dimensionless parameters $a_0$, $d_0$ ($c_3$, $d_5$ in the standard conventions). 
Given this starting point, we consider the interactions of the massive graviton with matter particles of spin 0, 1/2, and 1.  
At the level of on-shell 3-point amplitudes the coupling between matter and gravity is described by   \eref{CS_m3massless}. 
These are the same as in ordinary GR up to  the overall normalization parameter $c_X$. 
In other words, for each matter particle we allow the strength of its gravitational coupling to differ from the GR value $c_X=1$ that realizes the equivalence principle.  
The parameter space of the theory also includes the Wilson coefficients of the contact terms entering the 4-point and higher-point amplitudes.
These can be adjusted so as to regulate  the UV properties of scattering amplitudes.   

In this set-up, we calculated the tree-level  amplitudes for gravitational Compton scattering of matter, $\cM(XG \to XG)$.
For a generic point in the parameter space they grow with energy as $(E/\Lambda_3)^6$ for $E \gg m$, 
where $\Lambda_3$ is the strong coupling scale of graviton self-scattering amplitudes in pure dRGT.  
We found, however, that the UV behavior can be considerably softer in some regions of the parameter space. 
If and only if $c_X = 1$, then 
$\cM(XG \to XG) \sim (E/\tilde \Lambda_c)^4$ with $\tilde \Lambda_c = \sqrt{m \mpl}$  can be achieved after a judicious choice of the Wilson coefficients. 
In such a case the Compton amplitudes are much softer than the graviton self-scattering ones, and their onset of strong coupling is postponed to distance scales that are sub-milimeter for realistic graviton masses.   
One can further soften their UV behavior to 
$(E/\Lambda_c)^3$ where $\Lambda_c =  (m \mpl^2)^{1/3}$, in which case $\Lambda_c^{-1}$ is microscopic for realistic graviton masses. 
This is possible only for the special value of the dRGT parameter $a_0$, namely for $a_0 = 1$ ($c_3 = 1/6$).   
These conclusions are universal for all types of matter, independently of mass and spin. 

We also discussed the positivity bounds on the parameter space of our theory. 
Previous works showed that positivity restricts the $a_0$-$d_0$ parameter space of pure dRGT to a finite area,
while the beyond-positivity bounds force the dRGT cutoff to be orders of magnitude lower than $\Lambda_3$. 
In the case of matter-gravity couplings the impact of positivity is less dramatic. 
In the generic region of the parameter space the specific constraints are written down in \eref{DRGTs_posE6}, \eref{DRGTf_pos}, and  \eref{DRGTv_pos} for spin 0, 1/2, and 1 matter particles.  
One universal conclusion is that the parameter $c_X$ has to be positive for all types of matter, 
which of course allows for the special value $c_X = 1$ that realizes the equivalence principle.  
In the  region of the parameter space where Compton amplitudes are softer, $\cO(E^4)$ or $\cO(E^3)$, the positivity bounds are automatically satisfied. 
Finally, we discussed the beyond-positivity bounds, and we found that they provide another rationale for softening Compton amplitudes. 
In the generic region of the parameter space the beyond-positivity bounds are violated unless the Compton amplitudes calculated in our EFT are valid only up to 
$\Lambda \ll \Lambda_3$ such that $\Lambda^{-1}$ is an astronomical distance  scale.
The culprit here is the quick $\cO(E^6)$ growth of the amplitudes for matter scattering on the scalar polarization of the graviton.  
Thus, softer Compton amplitudes are necessary for our  matter-gravity interactions to emerge from a local, causal, and  Poincar\'{e} invariant UV completion above a reasonably high cutoff scale. Once the Compton amplitudes are softened down to $\cO(E^4)$, the beyond-positivity bounds are practically equivalent to the standard positivity ones, and they do not impose any additional constraints on the matter-gravity couplings.

\section*{Acknowledgments}

We thank Brando Bellazzini for valuable discussions and comments on the manuscript.  
AF is partially supported by the French Agence Nationale de la Recherche (ANR) under grant ANR-19-CE31-0012 (project MORA). 
\appendix

\providecommand{\href}[2]{#2}\begingroup\raggedright\endgroup

\end{document}

%% file: shortcuts.tex
\newcommand{\fref}[1]{Fig.~\ref{fig:#1}} 
\newcommand{\eref}[1]{Eq.~\eqref{eq:#1}}

\newcommand{\sref}[1]{Section~\ref{sec:#1}}
\newcommand{\cref}[1]{Chapter~\ref{ch:#1}}

\newcommand{\nnl}{\nonumber \\}

\newcommand{\beq}{\begin{equation}} 
\newcommand{\eeq}{\end{equation}} 
\newcommand{\ba}{\begin{array}}  
\newcommand{\ea}{\end{array}} 
\newcommand{\bea}{\begin{eqnarray}}  
\newcommand{\eea}{\end{eqnarray} }  
\newcommand{\be}{\begin{eqnarray}}  
\newcommand{\ee}{\end{eqnarray} }  
\newcommand{\bal}{\begin{align}}
\newcommand{\eal}{\end{align}}   
\newcommand{\bi}{\begin{itemize}}  
\newcommand{\ei}{\end{itemize}}  
\newcommand{\ben}{\begin{enumerate}}  
\newcommand{\een}{\end{enumerate}}  
\newcommand{\bc}{\begin{center}}
\newcommand{\ec}{\end{center}} 
\newcommand{\bt}{\begin{table}}
\newcommand{\et}{\end{table}}  
\newcommand{\btb}{\begin{tabular}}
\newcommand{\etb}{\end{tabular}}

\newcommand{\cO}{{\mathcal O}}

\newcommand{\cM}{{\mathcal M}}

\newcommand{\mpl}{M_{\mathrm Pl}}

\def\mpl{\, M_{\rm Pl}}

\newcommand{\eps}{\epsilon}
